# RIP: An Efficient Hybrid Repeater Insertion Scheme for Low Power

Xun Liu    Yuantao Peng
North Carolina State University
Raleigh, NC 27695
{xunliu,ypeng}@ncsu.edu

Marios C. Papaefthymiou
University of Michigan
Ann Arbor, MI 48109
marios@eecs.umich.edu

**Abstract**

*This paper presents a novel repeater insertion algorithm for interconnect power minimization. The novelty of our approach is in the judicious integration of an analytical solver and a dynamic programming based method. Specifically, the analytical solver chooses a concise repeater library and a small set of repeater location candidates such that the dynamic programming algorithm can be performed fast with little degradation of the solution quality. In comparison with previously reported repeater insertion schemes, within comparable runtimes, our approach achieves up to 37% higher power savings. Moreover, for the same design quality, our scheme attains a speedup of two orders of magnitude.*

## 1. Introduction

This paper presents a novel hybrid repeater insertion technique for low-power global interconnect designs. Given a two-pin interconnect and its timing budget, our algorithm derives the number of repeaters, and the width and location of each repeater so that the timing constraint is satisfied and power dissipation is minimized. The hybrid nature of our scheme stems from its judicious combination of an analytical repeater insertion solver and a dynamic programming (DP) based approach. Specifically, our algorithm proceeds in three steps. First, an initial repeater insertion solution is derived using DP with a very coarse repeater library. Second, an analytical procedure is applied to refine the initial solution and derive a new repeater library and a set of location candidates that fit the current design. Finally, the DP algorithm is repeated with the new library and location set for the low-power repeater insertion solution.

Our hybrid algorithm maintains the advantages of both analytical and DP-based schemes, producing high-quality interconnect designs efficiently. Furthermore, it is highly practical due to the adoption of a realistic interconnect model. Specifically, the interconnects are represented as a sequence of wire segments with fixed lengths and distinct RC characteristics, as derived from a routing procedure. Moreover, our algorithm can handle *forbidden zones*, i.e., parts of interconnects through macrocells in which no repeater can be placed, and is thus applicable to nets routed in real design scenarios.

We have implemented our repeater insertion algorithm into a software tool, called RIP, and applied it to the design of low-power global interconnects. In comparison with the conventional DP algorithms, our scheme achieves power reductions of up to 37% with comparable runtimes. Moreover, for the same design quality, RIP achieves shorter runtimes by two orders of magnitude.

The remainder of the paper is organized as follows. Section 2 describes previous research on repeater insertion. The problem of low-power repeater insertion for multi-layer two-pin interconnects is formulated in Section 3. In Section 4, analytical constraints on the repeater widths and locations are derived that must be satisfied to minimize repeater power dissipation. Our algorithm is presented in Section 5. Section 6 presents our experimental results. Section 7 summarizes our paper.

## 2. Previous Work on Repeater Insertion

Extensive work on repeater insertion has appeared in the literature [4, 19]. Repeater insertion has been applied to interconnect designs with various objectives such as delay minimization [3, 12, 17] and power minimization [5, 10, 13, 15, 16]. Several circuit models have been proposed to compute the delay and power dissipation of repeaters such as the switch-level RC model [8] and moment matching model [1]. In analytical repeater insertion schemes, the optimization objectives are described using analytical functions of repeater width and location. The optimal repeater insertion solutions can be derived by setting the derivatives of these functions to zero and solving the ensuing equations [6, 7]. Analytical schemes assume that the repeater widths and/or locations can be continuously changed. In actual designs, however, repeater widths are *discrete* due to layout design rules. In addition, the repeater locations are restricted to areas not occupied by circuit blocks. Consequently, when practical interconnects are considered, the analytical objective functions become very complex or even intractable.

To address the limitation of analytical schemes, a repeater insertion algorithm based on DP was proposed in [11] and improved in [20] for interconnect delay minimization. This algorithm has been modified for interconnect power reduction [14, 18, 21]. In these DP schemes, the possible widths and locations of the repeaters are discrete and finite. The algorithms choose the best solution out of all the possibilities. As a result, the efficiency of the DP schemes may be significantly affected by the given repeater library and potential repeater locations. Specifically, if the allowable repeater widths and locations are too limited, the quality of the interconnect may degrade substantially. On the other hand, when the numbers of repeater widths and locations are large, the DP algorithms become very time-consuming [14].

DP algorithms perform well for interconnect delay reduction, since the delay of a min-delay interconnect design is insensitive to its repeater widths and locations [9]. Consequently, a small-size coarse-granularity repeater library can be derived and applied to all interconnect designs with limited performance loss [2]. However, when power reduction is considered, i.e., the problem addressed in this paper, DP schemes become less effective. Power dissipation of repeaters is sensitive to repeater widths, since at a first order approximation, gate capacitance depends linearly on repeater widths. Consequently, fine granularities are needed for repeater widths, so that repeater widths close to the optima can be used. However, the runtime complexity of DP schemes becomes pseudo-polynomial when power minimization is considered [14]. The use of fine granularities leads to large numbers of possible repeater widths, resulting in excessively long runtimes.

## 3. Problem Formulation

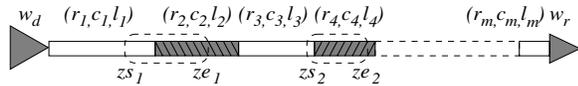

**Figure 1. Non-uniform two-pin interconnect.**

Figure 1 shows the structure of a multi-layer two-pin interconnect. The sizes of the driver and load are equal to $w_d$ and $w_r$, respectively. The interconnect is made of $m$ segments connected in a linear order as derived by a routing tool. The $i$th segment has a given length $l_i$, and the resistance and capacitance per unit length are $r_i$ and $c_i$, respectively. In a realistic interconnect routing scenario, the interconnect may go through some macro-blocks, in which no repeater can be placed. The portions of interconnect within the macro-blocks are represented as forbidden zones, whose ranges are labeled by $(zs_i, ze_i), i = 1, \ldots, b$.

The problem of *low-power repeater insertion for multi-layer two-pin interconnects* can be described as follows:

**Problem LPRI** Let $m$ be the number of interconnect segments, and let $l_i$, $r_i$ and $c_i$ be the length, resistance per unit length, and capacitance per unit length of the segment $i \in \{1, 2, \ldots, m\}$, respectively. Furthermore, let $b$ be the number of forbidden zones and $(zs_i, ze_i), i = 1, \ldots, b$ be the range of each zone $i$. Given the widths of the driver and receiver $(w_d, w_r)$ and a timing target $\tau_t$, compute the number of repeaters $n$ and the width $w_j$ and location $x_j$ of each repeater $j \in \{1, 2, \ldots, n\}$ such that $\forall j, x_j \notin (zs_i, ze_i), i = 1, \ldots, b$, the delay of the interconnect is equal to or less than $\tau_t$, and the total power $P$ of the repeaters is minimized.

## 4. Low-power Repeater Insertion Analysis
### 4.1. Repeater delay and power

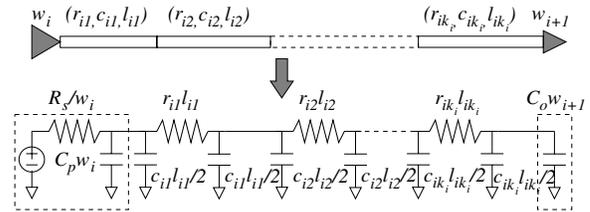

**Figure 2. Circuit model of a repeater stage.**

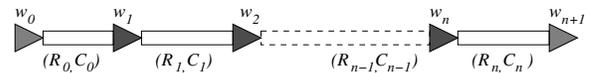

**Figure 3. Repeaters in a non-uniform net.**

In our analysis of repeater delays, we use the widely adopted Elmore delay model, although more accurate analytical delay models can be used by replacing the Elmore delay with the corresponding delay functions. Figure 2 shows the circuit model of a single repeater stage in which the driving repeater is represented using the switch-level RC model. Each interconnect segment is described using the lumped-RC $\pi$ model. The receiving repeater is modeled as a capacitor. The Elmore delay of such a stage is derived as:

$$\tau_i = R_s C_p + \frac{R_s}{w_i}(\sum_{j=1,k_i} l_{ij}c_{ij} + C_o w_{i+1}) + \sum_{j=1,k_i} l_{ij}r_{ij}C_o w_{i+1} + \sum_{j=1,k_i}(\frac{1}{2}l_{ij}c_{ij} + \sum_{h=j+1,k_i} l_{ih}c_{ih})r_{ij}l_{ij} , \quad (1)$$

where $R_s$, $C_o$, $C_p$ are the output resistance, input capacitance, and output capacitance of a unit-size repeater, $k_i$ is the number of segments between the repeaters $i$ and $i + 1$, and $(r_{ij}, c_{ij}, l_{ij})$ are the unit length resistance, unit length capacitance, and length of segment $j \in (1, 2, \ldots, k_i)$. When $n$ repeaters are inserted into a multi-layer two-pin net as shown in Figure 3, the total delay can be written as:

$$\tau_{total} = \sum_{i=0}^{n} \tau_i , \quad (2)$$

where $\tau_i$ is the Elmore delay of each stage. Note that, to simplify Equation (2), we denote the widths of the interconnect driver $w_d$ and receiver $w_r$ as $w_0$ and $w_{n+1}$, respectively.

From [5], the short-circuit power of repeaters is very small for advanced VLSI technologies. The total power of repeaters can therefore be approximated by the sum of the dynamic power and leakage power and is given by:

$$P = \alpha v_{dd}^2 f C_{total\_load} + \sum_{i=1}^{n} \beta w_i , \quad (3)$$

where $\alpha$ is the signal activities, $C_{total\_load}$ is the total gate capacitance, and $\beta$ is a constant. Since $C_{total\_load}$ is a linear function of the total repeater width, Equation (3) can be rewritten as follows:

$$P = c + \gamma \sum_{i=1}^{n} w_i , \quad (4)$$

where $c$ and $\gamma$ are constants. Consequently, the minimization of the repeater power dissipation is equivalent to the minimization of the total repeater width $p = \sum_{i=1}^{n} w_i$. Notice that the power dissipation due to interconnects is a constant and therefore not considered during our low-power repeater insertion derivation.

### 4.2. Constraints on repeater widths

In Problem **LPRI**, it can be proved that, when the power dissipation is minimized, we have

$$\tau_{total} = \tau_t . \quad (5)$$

Specifically, if $\tau_{total} < \tau_t$, the repeater widths can then be further decreased to reduce power. As a result, we can use Lagrange relaxation to analyze the optimal repeater insertion solutions. In particular, we define

$$L = p + \lambda(\tau_{total} - \tau_t) . \quad (6)$$

When power is minimized, we should have

$$\frac{\partial L}{\partial w_i} = 0 , \ i = 1, \ldots, n . \quad (7)$$

Combining Equations (2), (6), and (7), we have

$$1 + \lambda[C_o(R_{i-1} + \frac{R_s}{w_{i-1}}) - \frac{R_s(C_i + C_o w_{i+1})}{w_i^2}] = 0, i=1,\ldots,n, \quad (8)$$

where $R_{i-1} = \sum_{j=1,k_{i-1}} r_{(i-1)j} l_{(i-1)j}$ is the total interconnect resistance between repeaters $i-1$ and $i$, and $C_i = \sum_{j=1,k_i} c_{ij} l_{ij}$ is the total interconnect capacitance between repeaters $i$ and $i+1$, as shown in Figure 3.

### 4.3. Constraints on repeater locations

We next develop a set of constraints on the repeater locations. Specifically, we denote the locations of the repeaters as $x_i, i = 1, 2, \ldots n$. Since $\tau_{total}$ is a function of repeater widths $w_i$ and locations $x_i$, using the first order Lagrange expansion, we have

$$\Delta \tau_{total} = \sum_{i=1}^{n} \frac{\partial \tau_{total}}{\partial w_i} \Delta w_i + \sum_{i=1}^{n} \frac{\partial \tau_{total}}{\partial x_i} \Delta x_i . \quad (9)$$

Since $\tau_{total} = \tau_t$ for the optimal repeater insertion design,

$$\Delta \tau_{total} = 0 . \quad (10)$$

Combining Equations (9) and (10), we have

$$\sum_{i=1}^{n} \frac{\partial \tau_{total}}{\partial w_i} \Delta w_i + \sum_{i=1}^{n} \frac{\partial \tau_{total}}{\partial x_i} \Delta x_i = 0 . \quad (11)$$

From Equation (7), we have

$$1 + \lambda \frac{\partial \tau_{total}}{\partial w_i} = 0 \Rightarrow \frac{\partial \tau_{total}}{\partial w_i} = -\frac{1}{\lambda} . \quad (12)$$

Combining Equations (11) and (12), we have

$$\sum_{i=1}^{n} \Delta w_i = \lambda \sum_{i=1}^{n} \frac{\partial \tau_{total}}{\partial x_i} \Delta x_i . \quad (13)$$

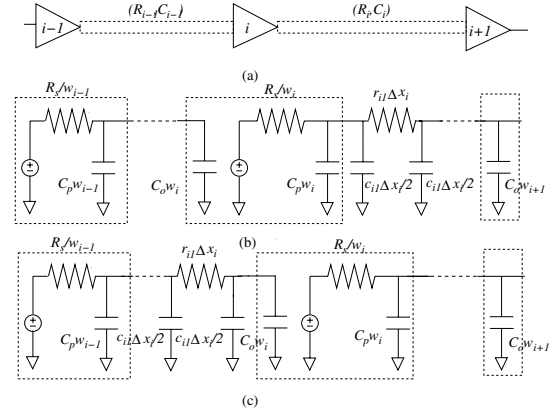

**Figure 4. Repeater downstream movement.**

The computation of $\partial \tau_{total}/\partial x_i$ can be described with the aid of Figure 4. When repeater $i$ moves downstream by a small distance $\Delta x_i$, a small resistive and capacitive load switches from the output to the input side of repeater $i$ as shown in Figure 4 (b) and (c), in which the circuits in the dotted boxes represent the repeaters. As a result, the delay increase between repeaters $i-1$ and $i$ can be calculated as

$$\Delta \tau_{i-1 \to i} = [C_o w_i r_{i1} + (R_{i-1} + \frac{R_s}{w_{i-1}}) c_{i1}] \Delta x_i . \quad (14)$$

Similarly, the delay increase between repeaters $i$ and $i+1$ is

$$\Delta \tau_{i \to i+1} = -[(C_i + C_o w_{i+1}) r_{i1} + \frac{R_s c_{i1}}{w_i}] \Delta x_i . \quad (15)$$

The total delay change due to the repeater movement can then be derived by combining Equations (14) and (15) as follows:

$$\Delta \tau_{total} = \Delta \tau_{i-1 \to i} + \Delta \tau_{i \to i+1} . \quad (16)$$

From Equation (16), we can derive the right-hand derivative of $\tau_{total}$ with respect to $x_i$ by letting $\Delta x_i$ approach zero:

$$(\frac{\partial \tau_{total}}{\partial x_i})_+ = \lim_{\Delta x_i \to 0+} \frac{\Delta \tau_{total}}{\Delta x_i} \quad (17)$$

$$= C_o r_{i1}(w_i - w_{i+1}) + R_s c_{i1}(\frac{1}{w_{i-1}} - \frac{1}{w_i}) + c_{i1} R_{i-1} - r_{i1} C_i.$$

Similarly, the left-hand derivative of $\tau_{total}$ with respect to $x_i$ is

$$(\frac{\partial \tau_{total}}{\partial x_i})_- = C_o r_{(i-1)k_{i-1}}(w_i - w_{i+1}) + R_s c_{(i-1)k_{i-1}}(\frac{1}{w_{i-1}} - \frac{1}{w_i})$$
$$+ c_{(i-1)k_{i-1}} R_{i-1} - r_{(i-1)k_{i-1}} C_i, \quad (18)$$

where $r_{(i-1)k_{i-1}}$ and $c_{(i-1)k_{i-1}}$ are the resistance and capacitance per unit length of the interconnect at the input of repeater $i$.

For the repeater insertion solution that minimizes power dissipation, the total repeater width cannot be further reduced, namely

$$\sum_{i=1}^{n} \Delta w_i \geq 0 . \quad (19)$$

Consequently, from Equation (13), we have

$$\lambda \sum_{i=1}^{n} \frac{\partial \tau_{total}}{\partial x_i} \Delta x_i \geq 0 . \quad (20)$$

Furthermore, since the movements of repeaters are independent from each other, we have

$$\forall i, \quad \lambda \frac{\partial \tau_{total}}{\partial x_i} \Delta x_i \geq 0 . \quad (21)$$

Therefore, we have

$$\lambda \left(\frac{\partial \tau_{total}}{\partial x_i}\right)_+ \geq 0 , \quad (22)$$

$$\lambda \left(\frac{\partial \tau_{total}}{\partial x_i}\right)_- \leq 0 . \quad (23)$$

When repeater $i$ is located inside an interconnect segment, we have $r_{i1} = r_{(i-1)k_{i-1}}$ and $c_{i1} = c_{(i-1)k_{i-1}}$. It follows that $\left(\frac{\partial \tau_{total}}{\partial x_i}\right)_+ = \left(\frac{\partial \tau_{total}}{\partial x_i}\right)_-$, and consequently, Inequalities (22) and (23) can be rewritten as

$$C_o r_{i1}(w_i - w_{i+1}) + R_s c_{i1}\left(\frac{1}{w_{i-1}} - \frac{1}{w_i}\right) + c_{i1}R_{i-1} - r_{i1}C_i = 0 . \quad (24)$$

## 5. Low-Power Repeater Insertion Algorithm

### 5.1. Analytical insertion

Solving the constraints in Section 4 for the optimal repeater widths and locations remains challenging, since $\tau_{total}$ is not an explicit analytical function of $x_i$. Furthermore, Relations (22) and (23) are inequalities. We next present algorithm REFINE, an iterative scheme that can derive a low-power repeater insertion solution from an initial solution.

Figure 5 shows the pseudocode of our algorithm REFINE. Given the specification of a multi-layer two-pin net and an initial repeater insertion solution, REFINE derives a new solution so that the total repeater width is minimized. Specifically, in Line 1, REFINE calculates the optimal repeater widths of the initial solution by solving the non-linear Equations (5) and (8) using Newton-Raphson method. The parameter $\lambda$ is also computed concurrently. It then calculates the total repeater width and initializes the loop-control variable $\varepsilon$ in Line 2. The optimization procedure is performed iteratively using the **while** loop in Lines 3–10. In Line 4, the left-hand and right-hand partial derivatives of $\tau_{total}$ with respect to the repeater location are computed for each repeater using Equations (17) and (18). If Relation (22) (or (23)) does not hold, repeater $i$ is moved downstream

REFINE
1 Compute $w_i$ and $\lambda$
2 $w_{total} = \sum_i w_i$ and $\varepsilon = \infty$
3 **while** $\varepsilon > \varepsilon_0$
4   Calculate $\left(\frac{\partial \tau_{total}}{\partial x_i}\right)_+$ and $\left(\frac{\partial \tau_{total}}{\partial x_i}\right)_-$ for each repeater
5   Move repeaters according to the results of Line 4
6   Update lumped $(R_i, C_i)$ driven by each repeater $i$
7   Solve for $w_i$ and $\lambda$
8   $w_{total\_old} = w_{total}, w_{total} = \sum_i w_i$
9   $\varepsilon = (w_{total\_old} - w_{total})/w_{total\_old}$
10 **return** $x_i, w_i$

**Figure 5. Algorithm** REFINE.

(or upstream) by a preselected distance. Based on Equation (13), such a movement results in the reduction of total repeater width. (If both relations do not hold, the moving direction is chosen for larger reduction.) A repeater will not be moved if the movement places the repeater inside a forbidden zone. After the movement, the lumped RC loads driven by each repeater are updated in Line 6. In Line 7, REFINE recalculates the width of each repeater and $\lambda$ using Equations (5) and (8). The total repeater width is then updated in Line 8. The iteration continues until the improvement is smaller than a preselected threshold $\varepsilon_0$. REFINE returns the repeater insertion solution in Line 10.

### 5.2. Repeater insertion heuristic

Though algorithm REFINE can quickly derive low-power repeater solutions, it assumes that the repeater width is continuously changeable, which is not practical. Furthermore, it requires an initial solution. In this section, we address these issues by combining REFINE with a DP algorithm.

RIP
1 Run DP with a coarse repeater library and locations
2 Run algorithm REFINE using the result from Line 1
3 Create a new library $B$ and a set of candidate locations $S$
4 Run DP with $B$ and $S$.

**Figure 6. Repeater Insertion Algorithm** RIP.

The pseudocode of our hybrid algorithm, called RIP, is given in Figure 6. RIP first performs a DP algorithm with coarse repeater widths and location candidates to derive an initial solution. Algorithm REFINE is then used to improve the initial solution in Line 2. In Line 3, RIP generates a new repeater library by rounding each repeater width from REFINE to its nearest valid discrete width. RIP also creates a small set of repeater location candidates by choosing several locations around the ones from REFINE. In Line 4, the DP algorithm is performed again with the new repeater library and location set to compute the final solution.

## 6. Experimental Evaluation

We have applied Algorithm RIP to various interconnect designs to demonstrate its effectiveness. Our interconnects and repeaters were generated using $0.18\mu m$ technology. Since our technique was used for global interconnects, each net was routed on metal4 and metal5 only. The number of segments for each net varied from 4 to 10. The length of each segment ranged from 1000 $\mu m$ to 2500 $\mu m$. A single forbidden zone was created for each net. The length of the zone ranged from 20% to 40% of the total interconnect length. The location of forbidden zone was uniformly distributed along the corresponding interconnect.

During the application of RIP, a DP algorithm was performed in conjunction with a library of 5 repeaters. The smallest repeater width and the width granularity were set to $80u$, where $u$ is the minimal repeater width. The candidate locations were uniformly distributed along the interconnects with a granularity of 200 $\mu m$, excluding the forbidden zone. The ensuing results were then improved using our analytical solver REFINE. The repeater widths and locations of the refined solution were used to generate a new repeater library and a new set of location candidates. In particular, we rounded each repeater size to the nearest width in a repeater library with a granularity of $10u$. The possible locations of repeaters included the locations derived by REFINE plus 10 locations before and after, with granularity 50 $\mu m$. Finally, the DP algorithm was performed using the new library and location set.

As a comparison, we used the DP algorithm in [14], one of the most prevalent and well cited schemes, to derive low-power repeater insertion solutions for the same interconnects and timing targets. We chose a repeater library of size 10 so that the total runtime of the DP scheme was comparable with that of our proposed scheme. We set the minimum repeater width to $10u$ and changed the granularity of the repeater widths. The candidate locations were uniformly distributed along the interconnects with a granularity of 200 $\mu m$, except for the forbidden zone.

Table 1 shows the experimental results from 20 interconnect designs. We designed each interconnect 20 times with timing targets ranging from 1.05 $\tau_{min}$ to 2.05 $\tau_{min}$, where $\tau_{min}$ is the minimum delay of the net. Columns 2–3 compare our scheme with the DP scheme in [14] with library size 10 and granularity $g = 10u$. As can be seen, our scheme can achieve maximal power savings $\Delta_{Max}$ up to 37.14%. Our scheme always succeeded in deriving solutions that satisfy the timing constraint, whereas the DP scheme resulted in several violations (6 out of 20 on the average), as shown in Column 3. When the library granularity $g$ increases to $20u$ and $40u$, the DP scheme can derive valid repeater insertion solutions for all the timing targets. As shown in Columns 5 and 7, our scheme achieves average power reductions $\Delta_{Mean}$ of 3.6% and 9.5% over the DP scheme, on the average.

| | g=10u | | g=20u | | g=40u | |
|---|---|---|---|---|---|---|
| Net | $\Delta_{Max}$ (%) | $V_{DP}$ | $\Delta_{Max}$ (%) | $\Delta_{Mean}$ (%) | $\Delta_{Max}$ (%) | $\Delta_{Mean}$ (%) |
| 1 | 22.95 | 7 | 10.00 | 4.72 | 28.57 | 11.41 |
| 2 | 17.39 | 6 | 13.33 | 2.57 | 15.79 | 7.11 |
| 3 | 26.19 | 6 | 7.69 | 2.85 | 21.05 | 8.47 |
| 4 | 15.87 | 3 | 20.00 | 2.60 | 21.43 | 8.35 |
| 5 | 20.69 | 9 | 13.33 | 4.83 | 21.43 | 9.96 |
| 6 | 15.69 | 4 | 15.38 | 3.27 | 28.57 | 9.98 |
| 7 | 15.69 | 7 | 10.53 | 2.27 | 28.57 | 8.96 |
| 8 | 30.19 | 7 | 20.00 | 5.97 | 30.00 | 12.19 |
| 9 | 20.00 | 7 | 10.00 | 4.27 | 28.57 | 11.31 |
| 10 | 13.89 | 6 | 7.69 | 2.78 | 21.05 | 8.37 |
| 11 | 23.40 | 6 | 12.50 | 2.12 | 30.00 | 9.43 |
| 12 | 10.96 | 6 | 7.69 | 3.14 | 20.00 | 7.44 |
| 13 | 17.86 | 8 | 10.00 | 3.95 | 21.43 | 8.54 |
| 14 | 29.82 | 9 | 9.09 | 3.35 | 28.57 | 10.20 |
| 15 | 19.23 | 5 | 10.00 | 2.65 | 21.43 | 8.32 |
| 16 | 37.14 | 8 | 17.11 | 5.53 | 21.43 | 11.62 |
| 17 | 24.62 | 6 | 9.09 | 3.19 | 26.67 | 9.84 |
| 18 | 4.35 | 4 | 8.33 | 2.47 | 21.43 | 8.41 |
| 19 | 11.11 | 7 | 9.09 | 3.36 | 21.43 | 9.80 |
| 20 | 29.58 | 7 | 16.13 | 5.50 | 21.43 | 10.90 |
| Ave | 20.33 | 6 | 11.8 | 3.6 | 23.94 | 9.53 |

**Table 1. Power reduction for two-pin nets.**

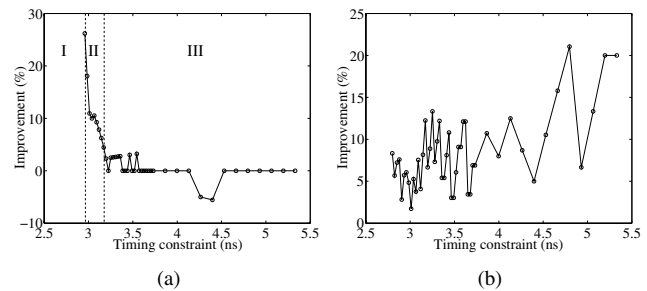

**Figure 7. Power savings over DP scheme [13] with repeater granularity (a)$10u$ (b)$40u$.**

Figure 7 reveals the relation between the timing targets and the power savings of our approach over the DP scheme with a repeater library of size 10. When the repeater granularity is small, large-size repeaters are missing in the library. As a result, the DP scheme fails to find any valid solution when the timing target is very tight, as shown by zone I in Figure 7(a). When the timing target is very loose, i.e., zone III, the DP scheme performs as well as our approach, since large-size repeaters are not required. In fact, there exist a limited number of cases when the DP scheme provides better solutions. Our scheme achieves the best power savings when the timing target is in zone II. On the other hand, if the repeater library has a coarse width granularity, our scheme always outperforms the DP scheme as shown in Figure 7(b). The power savings increase when the timing target becomes loose, since such interconnect designs require various small-size repeaters, which seldom exist in a coarse-granularity library.

To demonstrate the favorable runtime-quality tradeoff of our scheme, we compared Algorithm RIP with the DP scheme that used a repeater library with a fixed width range of ($10u$, $400u$). The repeater width granularity $g_{DP}$ varied from $40u$ to $10u$, resulting in different numbers of repeaters in the library. Table 2 shows that, when $g_{DP}$ decreases, the average power savings ($\Delta$) of our scheme over the DP scheme reduces from 14.2% to 0.3%. However, the runtime $T_{DP}$ increases significantly. The average runtime of our scheme is about 0.17 seconds, more than 200 times faster than the DP scheme when both schemes produce comparable results. It is worth mentioning that when $g_{DP} = 10u$, the DP scheme used the same repeater library as that of RIP. Still, Algorithm RIP achieved slightly better power savings than the DP scheme. Such a seemingly surprising result is due to the fact that RIP judiciously chooses the candidate repeater locations, which have a smaller local granularity than that of the DP scheme.

| $g_{DP}(u)$ | $\Delta(\%)$ | $T_{DP}$ (s) | Speedup |
|---|---|---|---|
| 40 | 14.2 | 1.0 | 6 |
| 30 | 7.8 | 1.9 | 11 |
| 20 | 4.0 | 5.8 | 34 |
| 10 | 0.3 | 34.45 | 203 |

**Table 2. Power savings and speedup tradeoff.**

## 7. Conclusion

In this paper, we propose a hybrid repeater insertion algorithm for interconnect power minimization. The novelty of our scheme is its judicious combination of an analytical solver with a DP scheme. Specifically, the analytical solver provides a concise repeater library and limited repeater location candidates, facilitating the efficient execution of the DP scheme. Our approach is highly practical, capable of handling multi-layer nets with forbidden zones. When applied to two-pin interconnects, our approach achieves up to 37% greater power savings or a speedup of more than 2 orders of magnitude in comparison with conventional DP schemes.

Our greedy analytical solver REFINE can be improved in several ways. Specifically, better power savings may be achieved if repeaters are allowed to move across small-size forbidden zones. Moreover, REFINE may be performed several times for further power reduction. We are currently extending our hybrid scheme to the design of low-power interconnect trees.

## References

[1] C. J. Alpert, A. Devgan, and S. T. Quay. Buffer insertion with accurate gate and interconnect delay computation. In *DAC*, June 1999.
[2] C. J. Alpert, R. G. Gandham, J. L. Neves, and S. T. Quay. Buffer library selection. In *ICCD*, Sept. 2000.
[3] C. J. Alpert, J. Hu, S. S. Sapatnekar, and P. G. Villarrubia. A practical methodology for early buffer and wire resource allocation. *IEEE Trans. CAD*, 22(5), May 2003.
[4] H. B. Bakoglu. *Circuits, Interconnects, and Packaging for VLSI*. Reading, MA: Addison-Wesley, 1990.
[5] K. Banerjee and A. Mehrotra. A power-optimal repeater insertion methodology for global interconnects in nanometer designs. *IEEE Trans. VLSI*, 49(11):2001–2007, Nov. 2002.
[6] C.-C. N. Chu and D. F. Wong. Closed form solution to simultaneous buffer insertion/sizing and wire sizing. In *Inter. Symp. on Physical Design*, Apr. 1997.
[7] C. C. N. Chu and D. F. Wong. A new approach to simultaneous buffer insertion and wire sizing. In *Inter. Conf. on CAD*, Nov. 1997.
[8] J. Cong, L. He, C. K. Koh, and P. H. Madden. Performance optimization of VLSI interconnect layout. *Integration, the VLSI Journal*, 21(1):1–94, Jan. 1996.
[9] J. Cong, T. Kong, and D. Z. Pan. Buffer block planning for interconnect-Driven floorplanning. In *Inter. Conf. on CAD*, Nov. 1999.
[10] G. S. Garcea, N. P. van der Meijs, and R. H. Otten. Simultaneous analytical area and power optimization for repeater insertion. In *Inter. Conf. on CAD*, Nov. 2003.
[11] L. Ginneken. Buffer placement in distributed RC-tree networks for minimal Elmore delay. In *Inter. Symp. on Circuits and Systems*, 1990.
[12] N. Hedenstierna and K. O. Jeppson. CMOS circuit speed and buffer optimization. *IEEE Trans. CAD*, 6(2):270–280, Feb. 1987.
[13] P. Kapur, G. Chandra, and K. C. Saraswat. Power estimation in global interconnect and its reduction using a novel repeater optimization methodology. In *DAC*, June 2002.
[14] J. Lillis, C. K. Cheng, and T.-T. Y. Lin. Optimal wire sizing and buffer insertion for low power and a generalized delay model. *J. of Solid-State Circuits*, 31(3):437–447, Mar. 1996.
[15] X. Liu, Y. Peng, and M. C. Papaefthymiou. Practical repeater insertion for low power: What repeater library do we need? In *DAC*, June 2004.
[16] A. Nalamalpu and W. P. Burleson. A practical approach to DSM repeater insertion: Satisfying delay constraints while minimizing area and power. In *IEEE International ASIC/SOC Conference*, Sept. 2001.
[17] M. Nekili and Y. Savaria. Optimal methods of driving interconnections in VLSI circuits. In *Inter. Symp. on Circuits and Systems*, May 1993.
[18] T. Okamoto and J. Cong. Buffered Steiner tree construction with wire sizing for interconnect layout optimization. In *Inter. Conf. on CAD*, 1996.
[19] R. Otten. Global wires harmful? In *Inter. Symp. on Physical Design*, Apr. 1998.
[20] W. Shi and Z. Li. An O(nlogn) time algorithm for optimal buffer insertion. In *DAC*, June 2003.
[21] H. Zhou, D. F. Wong, I.-M. Liu, and A. Aziz. Simultaneous routing and buffer insertion with restrictions on buffer locations. *IEEE Trans. CAD*, 19(7):819–824, 2000.